\documentclass[twocolumn,prl,superscriptaddress,showpacs]{revtex4}
\usepackage{graphicx}
\usepackage{pslatex}
\usepackage{amsmath,amssymb}

\newcommand{\bra}[1]{\left\langle{#1}\right|}
\newcommand{\ket}[1]{\left|{#1}\right\rangle}

\begin{document}

\title{Local interactions and non-Abelian quantum loop gases}

\author{Matthias Troyer}
\affiliation{Theoretische Physik, Eidgen\"ossische Technische
Hochschule Z\"urich, 8093 Z\"urich, Switzerland}
\author{Simon Trebst}
\affiliation{Microsoft Research, Station Q, University of California,
Santa Barbara, CA 93106}
\author{Kirill Shtengel}
\affiliation{Department of Physics and Astronomy, University of
California, Riverside, CA 92521}
\author{Chetan Nayak}
\affiliation{Microsoft Research, Station Q, University of California,
Santa Barbara, CA 93106}

\date{\today}

\begin{abstract}
Two-dimensional quantum loop gases are elementary examples 
of topological ground states with Abelian or non-Abelian anyonic excitations. 
While Abelian loop gases appear as ground states of local, gapped 
Hamiltonians such as the toric code, we show that gapped non-Abelian loop gases 
require non-local interactions (or non-trivial inner products). 
Perturbing a local, gapless Hamiltonian with an anticipated ``non-Abelian'' 
ground-state wavefunction immediately drives the system into the Abelian 
phase, as can be seen by measuring the Hausdorff dimension of loops.
Local quantum critical behavior is found in a loop gas in which all 
equal-time correlations of local operators decay exponentially. 
\end{abstract}



\maketitle


\noindent
Non-Abelian topological phases are the focus
of considerable excitement as a result of their universality,
their novelty, their beautiful mathematical properties, and their potential
application to quantum computing \cite{Nayak08}. 
However, the only concrete physical system in which there is any 
experimental evidence for a topological phase is the 
two-dimensional electron gas at high magnetic fields, i.e. in the 
quantum Hall regime.
In order to find such phases elsewhere -- in transition metal oxides
or in ultra-cold atomic gases -- it is important for
theory to serve as a guide by identifying conditions which a
system must satisfy in order to support a non-Abelian
topological phase.

One simple class of models is associated with {\it quantum loop
gases} (QLG), in which an orthonormal basis of the low-energy Hilbert space
can be mapped onto configurations of loops \cite{Freedman04a}.
One remarkable feature of topological phases
is that the ground-state wavefunction encodes many of
the quasiparticle properties \cite{Freedman04a,Kitaev06b,Levin06a},
which was exploited as far back as Laughlin's pioneering work on the 
fractional quantum Hall effect \cite{Laughlin83}. Therefore, the 
ground-state wavefunction plays a central role in the theory. 
Many of the properties of a QLG can be deduced by mapping 
the ground state to a classical statistical mechanical model.

The {\it toric code} \cite{Kitaev03} is the classic example
of a QLG; the associated classical model is critical percolation.
However, the toric code Hamiltonian is in a
$\mathbb{Z}_2$ topological phase which is Abelian; i.e. all of its
quasiparticles are Abelian anyons. As we discuss below,
non-Abelian topological ground states should be associated with
critical O($n$) loop models with $n>1$ \cite{Freedman04a,Freedman05a}.
(The $n=1$ case is equivalent to critical percolation.) 
However, local gapped Hamiltonians with these ground states are 
not known. In this paper, we show that such Hamiltonians do not exist.
Hamiltonians with the desired ground states have been constructed in
Refs.~\onlinecite{Freedman05a,Freedman05b}, but these models 
are gapless and describe critical points, not stable phases.
It was conjectured that by perturbing such a critical model, one could drive 
the system into a gapped non-Abelian topological phase. 
In this paper, we analyze the instabilities of such critical models 
and show that perturbations fundamentally alter the nature of the ground 
state. For instance, one of the simplest relevant perturbations drives the system 
into the {\em Abelian} $\mathbb{Z}_2$ topological phase.
Therefore, non-Abelian topological phases require more intricate
Hamiltonians.

\paragraph{Loop gas wavefunction.--}
QLGs can be realized in lattice models whose low-energy Hilbert space is spanned by
states of the form $|\mathcal{L} \rangle$, where the multi-loop $\mathcal{L}$ is an arbitrary
collection of non-intersecting loops.
Here we consider the `$d$-isotopy' wavefunction which generalizes the 
ground state (GS) of the toric code \cite{Freedman05a}:
\begin{equation}
\left|\Psi_0^{(d)} \rangle\right. \propto
\sum_{\{\mathcal{L}\}}d^{\ell(\mathcal{L})}
\left|\mathcal{L} \rangle\right. .
 \label{eq:d-iso-GS}
\end{equation}
In this expression, $\ell(\mathcal{L})$ is the number of loops
in the multi-loop $\mathcal{L}$. The loops have a `fugacity' $d$.
In the toric code, $d=1$. 
This wavefunction has two key features which hint at its
topological nature: On non-trivial surfaces, there is
a (degenerate) space of such wavefunctions corresponding,
for instance, to different winding numbers.
Second, the wavefunction amplitude
is independent of the lengths of the loops.
However, the latter is neither necessary nor sufficient
for a gapped topological phase \cite{Freedman05b}.

In a topological phase, the parameter $d$ also determines the topological 
properties of excitations, as discussed in Ref. \onlinecite{Freedman04a}.
Excitations can be studied by considering the ground state on a surface
with punctures: each puncture can be viewed as a localized excitation
which is specified by the boundary conditions at the puncture. If loops
terminate at the boundary, the excitation is non-trivial (this is sufficient but
not necessary). The amplitude to create a pair of such quasiparticles
and annihilate them later is a measure of the number of states of
such a quasiparticle called the {\it quantum dimension}.
If there are $N$ quasiparticles with quantum dimension $D$,
there will be $\sim D^N$ degenerate states.
For $D>1$ and not an integer, there will be a large degeneracy which
cannot be ascribed locally to the quasiparticles, so they will have 
non-Abelian braiding statistics.
The universal properties of a topological phase are independent
of any coordinate system; in particular, space and time can be interchanged.
Therefore, the quantum dimension can be determined directly from the
ground-state wavefunction. For a topological phase with ground state
(\ref{eq:d-iso-GS}), the quantum dimension of the fundamental quasiparticle is 
equivalent to the fugacity $d$.
A loop can be viewed as the projection onto a fixed time slice
of a pair creation and annihilation process. For $d=\sqrt{2}$, the
fundamental quasiparticle has the same quantum dimension
as the $\sigma$-field in the Ising topological quantum field theory
or the spin-$1/2$ field in SU(2)$_2$.
For arbitrary fugacity $d$, the loop gas ground state (\ref{eq:d-iso-GS})
has no relation to any known topological phase.

The fugacity $d$ also determines the correlation functions of the
associated classical statistical model, which is the O($n$) loop
model with $n=d^2$:
\begin{equation}
\label{eqn:O(n)-def}
Z_{\rm O(n)}(x) \equiv {\sum_\alpha} \left(\frac{x}{n}\right)^{b(\mathcal{L})}\,
{n^{\ell(\mathcal{L})}} \,.
\end{equation}
Here, $b(\mathcal{L})$ is the total length of the multi-loop $\mathcal{L}$.
For integer $n$, the right-hand-side is the expansion in
powers of $x$ of the Boltzmann weight
$e^{-\beta H} = \prod_{\langle i,j\rangle} (1+x{\hat{S}_i}\cdot {\hat{S}_j})$
for a model of classical interacting spins with O($n$) symmetry.
For $x=n$, 
$Z_{\rm O(n)}(x)=\langle {\Psi_0^{(d)}}|{\Psi_0^{(d)}} \rangle$,
so the equal-time ground-state correlations
contained in the QLG's $|{\Psi_0^{(d)}} \rangle$ can be obtained from
the known correlations of $Z_{\rm O(n)}(x)$.
For $n\leq2$ and $x\geq{x_c}=n/\sqrt{2+\sqrt{2-n}}$,
this model is in its low-temperature phase,
which is critical \cite{Nienhuis87}. 
It is necessary for the loops to be critical in order for $|{\Psi_0^{(d)}}\rangle$ 
to be the ground state of a topological phase; only then will the endpoints 
of a broken loop be deconfined.
Therefore, for $|{\Psi_0^{(d)}}\rangle$ to be in a topological
phase it is required that $d \leq \sqrt{2}$.
However, it is equally important that correlation functions of local
operators decay exponentially in time 
since a topological phase requires a gap to excited states.

\paragraph{Hamiltonian.--}

\begin{figure}[t]
\includegraphics[width=\columnwidth]{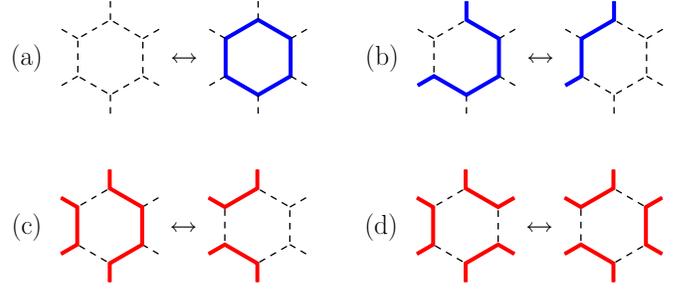}
\caption{Four types of moves enacted by the plaquette flip term $F_p$.
 Term (a) enforces the loop fugacity $d$, (b) enforces isotopy invariance.
 Collectively they are known as `$d$-isotopy' moves.
 Terms (c) and (d) are the surgery terms appropriate for $d=1$.
 }
\label{fig:moves}
\end{figure}

The wavefunction (\ref{eq:d-iso-GS}) is the ground state of the following spin-1/2 Hamiltonian,
where the spins live on the edges of a honeycomb lattice:
\begin{multline}
  \mathcal{H}_0^{(d)}= J{\sum_v} \biggl(1+{\prod_{i\in e(v)}}{\sigma^z_i}\biggr)
\\
  + \frac{K}{2} \sum_{p} \biggl[\frac{2}{1+d^2}
    {\left({d\openone - F_p}\right)\mathbb{P}^0_p 
     \left({d\openone - F_p}\right)} + 
    {\left({\openone - F_p}\right)\mathbb{P}^1_p}\biggr] \,.
 \label{eq:Hamiltonian0}
\end{multline}
The dominant first term ($J\gg K$) enforces the constraint that the low energy Hilbert
space is spanned by configurations with an even number of ${\sigma^z}=1$
spins on edges $e(v)$ around all vertices $v$.
The loops are formed by the ${\sigma^z}=1$ edges.
${F_p} \equiv \prod_{i\in p} \sigma^x_i$ flips the six spins around a hexagonal
plaquette $p$ and $\mathbb{P}^m_p$ are projectors onto
configurations with $m$ loop segments around a given hexagon. (Here we
define $\mathbb{P}^1_p$ so that it annihilates states with a single loop
forming the plaquette boundary as in Fig.~\ref{fig:moves}(a); only
configurations of the type shown in Fig.~\ref{fig:moves}(b) are not
annihilated by $\mathbb{P}^1_p$.) 
Notice that the Hamiltonian~(\ref{eq:Hamiltonian0}) includes processes shown
in Fig.~\ref{fig:moves}(a,b), but  \emph{not} those in Fig.~\ref{fig:moves}(c,d). 
The most salient property of this
Hamiltonian is that it is a sum of projectors which simultaneously
annihilate the wavefunction (\ref{eq:d-iso-GS}) which is, therefore,
the ground state. The low-energy spectrum of the
Hamiltonian~(\ref{eq:Hamiltonian0}) is gapless \cite{Freedman05b}
for $d\leq\sqrt{2}$: the quantum dynamics  is very inefficient at mixing states 
with different loop sizes, thereby resulting in gapless modes.

For $d=1$ a gap to all excited states can be opened {\it without} changing the GS $|\Psi_0^{(1)}\rangle$ 
by augmenting (\ref{eq:Hamiltonian0})  with the ``loop surgery'' terms in 
Fig.~\ref{fig:moves}(c,d)
\begin{equation}
  \mathcal{H}_{\text{TC}}=
 \mathcal{H}_0^{(d=1)}  + \frac{K}{2} \sum_p \biggl[\left({\openone -
F_p}\right)\left(\mathbb{P}^2_p+ \mathbb{P}^3_p\right)\biggr] \,.
 \label{eq:HamiltonianTC}
\end{equation}
This is the honeycomb lattice version of the
toric code Hamiltonian \cite{Kitaev03}, hence its GS is in a $\mathbb{Z}_2$
topological phase. 
Clearly, the existence of an energy gap above a ground state depends on
the Hamiltonian.

Augmenting the Hamiltonian~(\ref{eq:Hamiltonian0}) by the surgery terms of 
Fig.~\ref{fig:moves}(c,d) for $d\neq 1$ is not straightforward, as these
terms generically do not conserve the number of loops and hence
skew the loop amplitudes in Eq.~(\ref{eq:d-iso-GS}). To preserve the correct
amplitudes, one may propose a Hamiltonian
\begin{equation}
  \mathcal{H}_{1}=
 \mathcal{H}_0^{(d)}  + \frac{K}{2} \sum_p \biggl[\left({\openone -
d^{\Delta l} F_p}\right)\left(\mathbb{P}^2_p+ \mathbb{P}^3_p\right)\biggr],
 \label{eq:Hamiltonian1}
\end{equation}
where $\Delta l$ is the change in the number of loops when plaquette
$p$ is flipped. Although our Monte Carlo simulations show that this
opens a gap $\Delta\approx 2K$, the problem is that $\Delta l$ is a {\it non-local} 
operator: its eigenvalue depends on how the loop segments are connected away 
from a given plaquette $p$.

\paragraph{Local Perturbations.--}
The question that we now want to address is: Can the non-Abelian state 
$|\Psi_0^{(d)}\rangle$ with $d\neq 1$ be the GS of a {\it local}, gapped 
Hamiltonian or is this specific to the Abelian case $d=1$? 

One proposal is adding a \emph{local} $(k+1)$-strand surgery term (not shown)
given by a lattice version of the Jones-Wenzl projector 
\cite{Freedman03,Freedman04a} which also annihilates the
GS $|\Psi_0^{(d)}\rangle$ if $d=2\cos\pi/(k+2)$.
However, most of the states in this sequence $d=1,\sqrt{2},(1+\sqrt{5})/2,\ldots$
($k=1,2,3,\ldots$) occur at values of $d$ for which the loops are not critical
\cite{Nienhuis87}.
The case $d=1$ ($k=1$) is the surgery term discussed above, and therefore we 
focus on $d=\sqrt{2}$ ($k=2$).
It can be argued that this term will \emph{not} open a gap, because
there is a vanishing probability for three {\it long} strands to meet
\cite{Schramm-private}.

\begin{figure}[t]
  \includegraphics[width=\columnwidth]{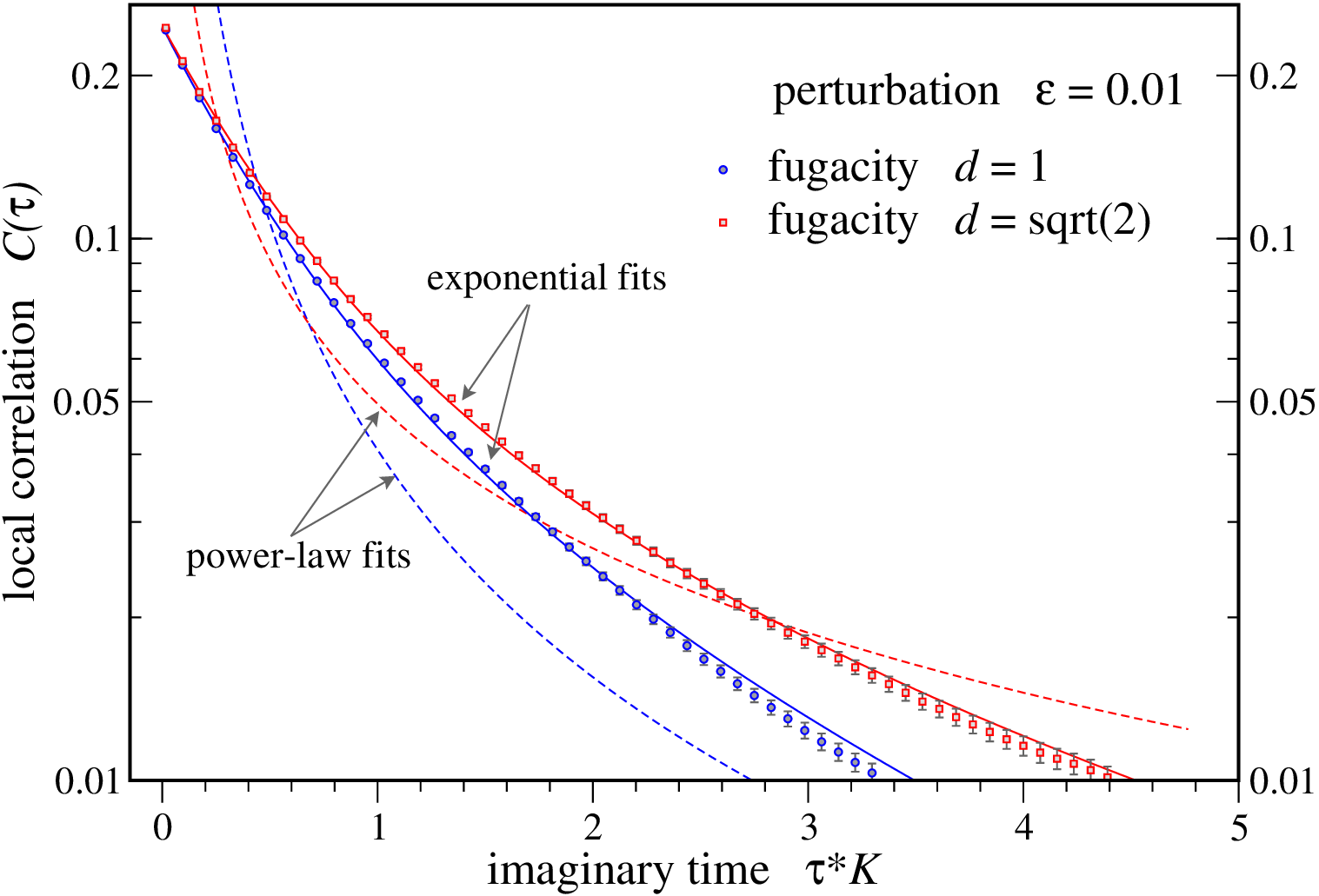}
  \caption{
    (color online) The local imaginary time correlation function $C(\tau)$ for
     the loop gas with fugacity $d$ and surgery term $\epsilon=0.01$. 
     Data shown are for system size $L=16$ and inverse temperature
     $\beta=16$.  }
  \label{Fig:CorrelationGap}
\end{figure}

However, it is possible for two strands to meet, so we consider the effect
of the {\it local} 2-strand surgery term:
\begin{equation}
  \mathcal{H}_{2}=
 \mathcal{H}_0^{(d=\sqrt{2})}  + \frac{\epsilon}{2} \sum_p \biggl[\left({\openone -
 F_p}\right)\left(\mathbb{P}^2_p+ \mathbb{P}^3_p\right)\biggr].
 \label{eq:Hamiltonian2}
\end{equation}
The state~$|{\Psi_0^{(\sqrt{2})}}\rangle$ is no longer the GS of Hamiltonian~(\ref{eq:Hamiltonian2}), 
because the second term is the surgery term for $d=1$.
Nevertheless, one may hope that for $\epsilon \ll K$, a weakly-perturbed, \emph {gapped} GS will emerge which might still be in the desired non-Abelian topological phase. 
We simulate the model~(\ref{eq:Hamiltonian2}) on a torus of $L \times L$ plaquettes with
tilt angle $60^{\circ}$ using a variant of the path-integral ground-state (PIGS)  algorithm \cite{Sarsa00}. 
We measure ground-state expectation values $\bra{0} A \ket{0}$ by sampling the 
continuous-time path integral representation of
\begin{equation}
\langle0| A |0\rangle = \lim_{\beta\to\infty} \frac{\langle \Psi_0^{(\sqrt{2})} |e^{-\beta H/2} A e^{-\beta H/2}|\Psi_0^{(\sqrt{2})}\rangle}{\bra{\Psi_0^{(\sqrt{2})}} e^{-\beta H}  \ket{\Psi_0^{(\sqrt{2})}}}
\end{equation}
using local updates. We choose the projection time $\beta$ large enough to project out all the states but the ground state. 
We find that this perturbation 
immediately opens a gap, since the time dependent  local correlation functions
\begin{eqnarray}
C(\tau) &=& \langle 0| \sigma_i^z e^{-\tau (H-E_0)}\sigma_i^z |0\rangle  \propto \\
&& \lim_{\beta\to\infty} \langle \Psi_0^{(d)} |e^{-(\beta-\tau) H/2} \sigma_i^z  e^{-\tau H}\sigma_i^z  e^{-(\beta-\tau) H/2}|\Psi_0^{(d)}\rangle, \nonumber
\end{eqnarray}
decrease exponentially as $C(\tau) \sim A(\tau) \exp(-\tau\Delta)$, where the prefactor 
$A(\tau) = (1-\exp(-\tau a k^2)/a\tau$
is determined by a quadratic dispersion $E(k) = \Delta + a k^2$
above the gap $\Delta$.
The 2-strand surgery term is expected to be a relevant perturbation, and an infinitesimal $\epsilon$
should be sufficient to open a gap. As shown in Fig. \ref{Fig:CorrelationGap}, a substantial gap exists already for a small $\epsilon=0.01$.

\begin{figure}[t]
 \includegraphics[width=\columnwidth]{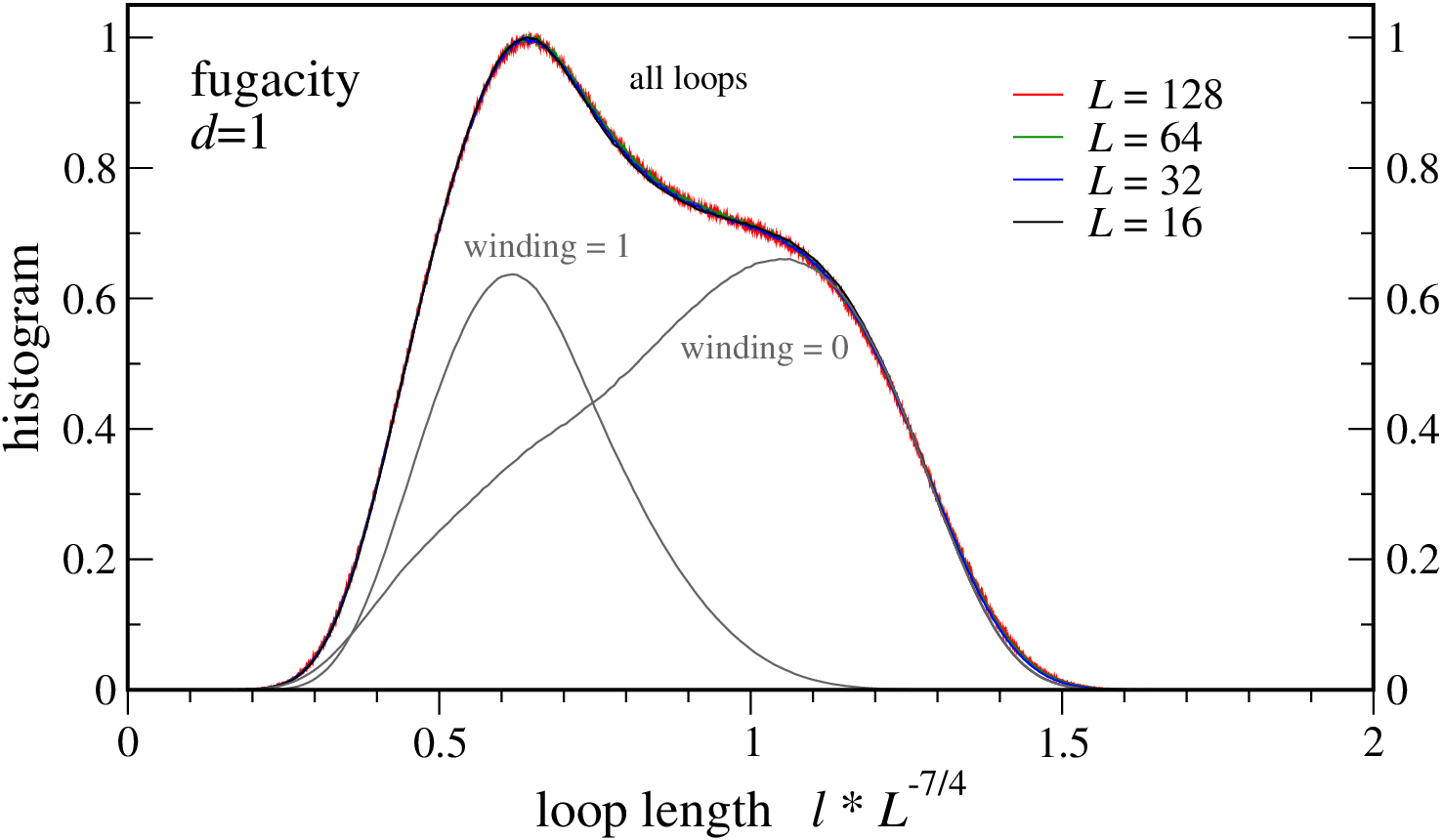}
 \vskip 2mm
 \includegraphics[width=\columnwidth]{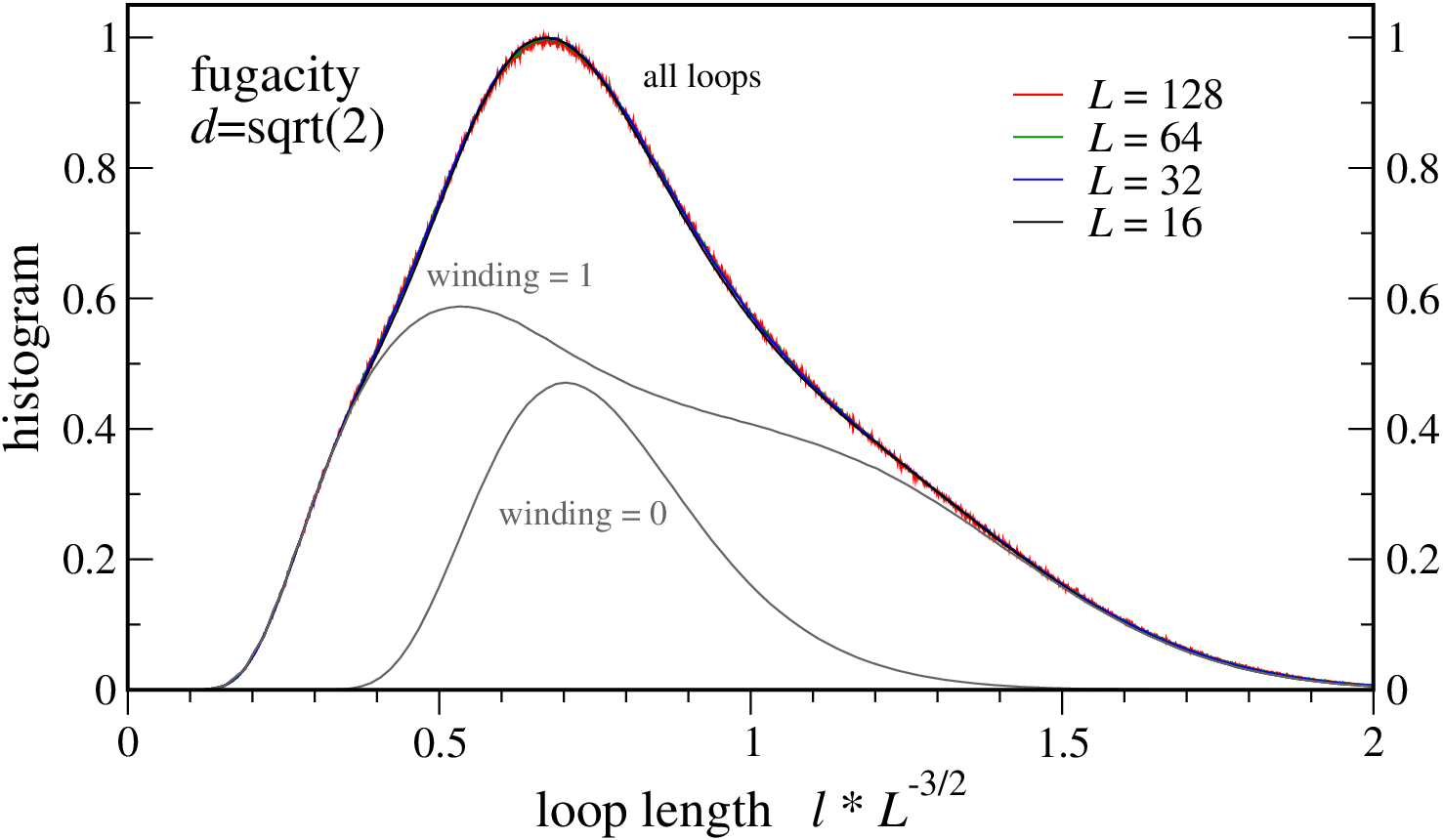}
 \caption{
   (color online) 
   Length distribution of the longest loop in the Abelian (non-Abelian)
   quantum loop gas with loop fugacity $d=1$ ($d=\sqrt{2}$).
   The multi-peak structure emerges from loops with different winding numbers.
   The data for various system sizes are rescaled by $L^{d_H}$, with the 
   Hausdorff dimension $d_H=7/4$ ($d_H=3/2$) in the Abelian (non-Abelian) case.
     }
 \label{Fig:HaussdorfDimension}
\end{figure}

We now need to determine whether this gapped ground state is in an Abelian 
or non-Abelian phase.
While these two phases can, in principle, be distinguished by the fugacity
of large loops, this is hard to measure. A simpler and more direct measurement 
differentiating the two phases is the Hausdorff dimension $d_H$ of the longest
loop. 
The Coulomb gas solution \cite{Nienhuis87} of the O($n$) loop model
allows to calculate $d_H$ exactly \cite{Saleur87} with $d_H=7/4$ for $|{\Psi_0^{(1)}}\rangle$ and $d_H=3/2$ for $|{\Psi_0^{\sqrt{2})}}\rangle$.
We first measure these Hausdorff dimensions 
by sampling multi-loops $\mathcal{L}$ in a Monte Carlo simulation with weights
given by the amplitude $d^{2\ell(\mathcal{L})}$. 
In Fig.~\ref{Fig:HaussdorfDimension} we plot the length distributions of the 
longest loops for $d=1$ and $d=\sqrt{2}$. Rescaling the data for various
system sizes by the expected Hausdorff dimension $L^{d_H}$ we find an 
excellent data collapse. The characteristic multi-peak structure of the distributions
originates from loops with different winding numbers 
\footnote{
While our Monte Carlo simulation conserves the parity of the total winding number 
of the multi-loop to be even, individual loops with odd winding exist.
}.

We have explicitly checked that the Hausdorff dimension is universal and constant 
for the full extent of a gapped phase by sampling the loop configurations created 
by domain walls of a ferromagnetic Ising model in its high temperature phase.  
At infinite temperature the Ising model on the triangular lattice is just percolation at 
the critical point, and the boundaries of percolation clusters form a loop gas  $|{\Psi_0^{(1)}}\rangle$. 
Although at any finite temperature above the critical point, corresponding to different 
GS wavefunctions within the same topological phase, the fugacity for small loops is 
changed, we find that the Hausdorff dimension of the longest 
loop stays $d_H=7/4$.

\begin{figure}[t]
  \includegraphics[width=\columnwidth]{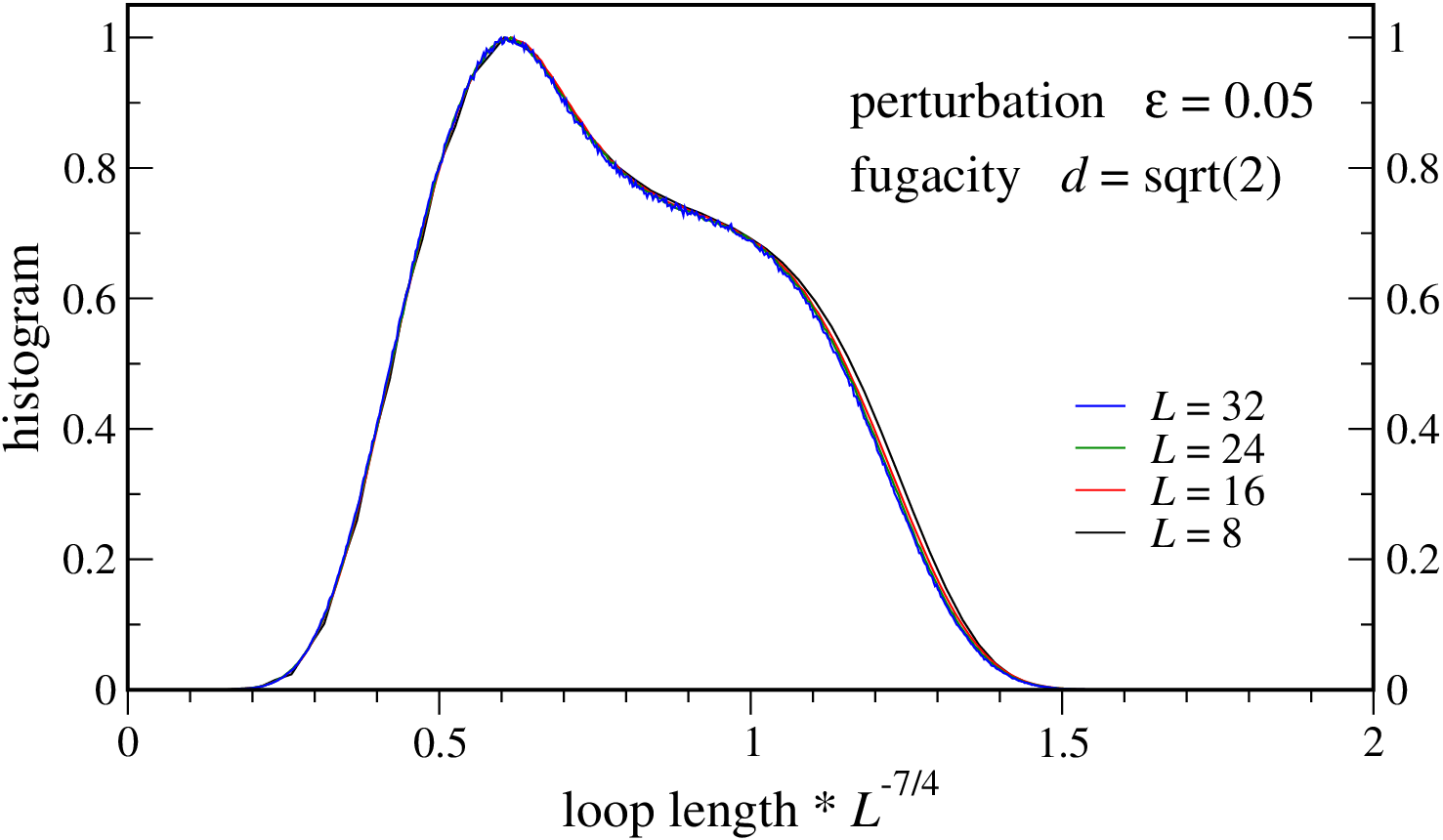}
  \caption{
    (color online) Data collapse of the rescaled loop length distributions 
    of various system sizes for small surgery term $\epsilon=0.05$.
    The loop fugacity is $d=\sqrt{2}$, and inverse temperature $\beta=16$. 
   }
  \label{Fig:LoopCollapse}
\end{figure}

Returning to model~(\ref{eq:Hamiltonian2}), we now analyze the Hausdorff dimension of the 
ground state in the gapped phase for small $\epsilon$. As shown in Fig. \ref{Fig:LoopCollapse} 
the perturbation changes the Hausdorff dimension from $d_H=3/2$ to $d_H=7/4$ and the
characteristic loop distribution to that of the {\it Abelian} phase.
This can be understood in a renormalization-group sense by noting that $\mathcal{H}_0^{(d)}$ enforces the loop fugacity $d$ only on microscopic loops, while the perturbation (which is the surgery term of the $d=1$ loop gas) acts on large loops, and is thus a {\em relevant} perturbation, driving the system to the Abelian $d=1$ fixed point.

\begin{figure}[t]
  \includegraphics[width=\columnwidth]{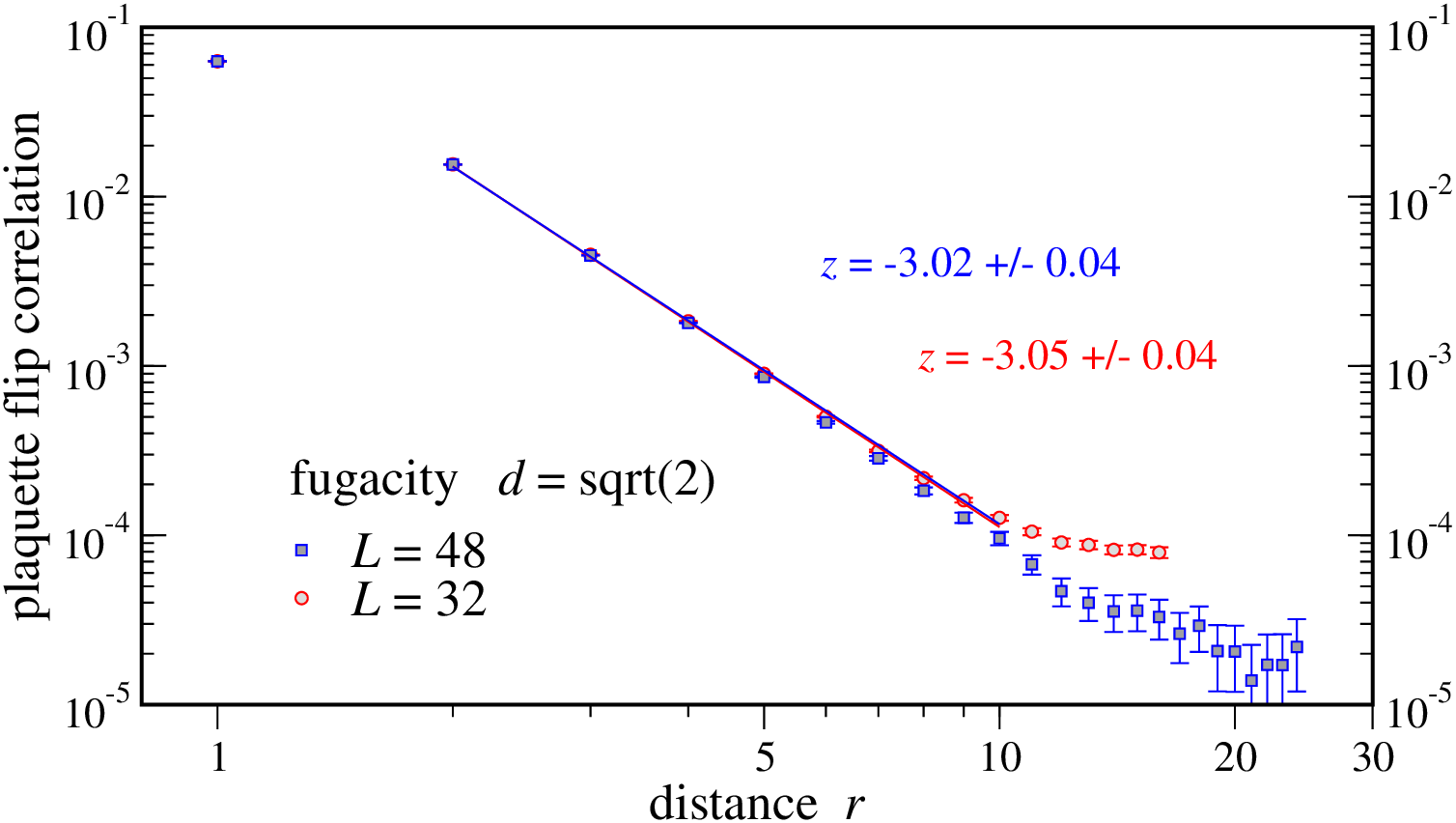}
  \caption{
    (color online) Algebraic decay of the plaquette flip correlations between plaquettes 
    at distance $r$ in the QLG $|{\Psi_0^{(\sqrt{2})}}\rangle$.
   }
  \label{Fig:CorrelationFlips}
\end{figure}

One might still hope that a different local perturbation to the Hamiltonian might open a gap but leave the system in the non-Abelian phase. We will now show that such hope is futile by showing that they are power-law correlation between {\it local} operators in the GS $|{\Psi_0^{(d)}}\rangle$ for 
$1< d\leq \sqrt{2}$.
While the correlations of most local operators are short-ranged, those of the plaquette flip operator 
$F_p$ decay algebraically as $r^{-z}$ with $z\approx 3$ for $d=\sqrt{2}$
as shown in Fig. \ref{Fig:CorrelationFlips} and smaller values of $z$ for $d<\sqrt{2}$.
A theorem by Hastings \cite{Hastings04a} then proves that because
of this algebraic (and not exponential) decay of the correlation function
between two local operators, the $d\ne1$ loop gas wavefunction
cannot be the ground state of a gapped local Hamiltonian.

The origin of the algebraic decay is the fractal nature of the loop gas: plaquette flips performing surgery between two segments of the same loop are correlated, since the change in loop number $\Delta l$ is not just the sum of the changes of the individual flips. Since the flip matrix element is $d^{\Delta l}$ this results in an algebraically decaying correlation function for $d\ne1$. This argument  applies not only to the wavefunction (\ref{eq:d-iso-GS}) but to any other wavefunction with long critical loops and a fugacity $d\ne1$ for large loops. 
Since the surgery term requires at least two loops to pass through the same plaquette or, in other words, four loop segments emanating from the same plaquette, the $\langle{F_p}(r)\,{F_p}(0)\rangle$ correlation function should not decay faster than the O($n$) model exponent associated with four-line vertices which gives $z\leq4$ \cite{Nienhuis87}. 

We can thus prove that there exists no local, gapped Hamiltonian whose ground-state wavefunction is a loop gas with non-Abelian excitations.
To construct local Hamiltonians with non-Abelian excitations one needs to go beyond loop gases and consider more intricate Hamiltonians, such as  string-net models \cite{Levin05a,Fidkowski07a}. 
They can be constructed systematically from loop gases with modified inner products
\cite{Fendley08}.

For $d=1$ the Hamiltonian (\ref{eq:Hamiltonian0}) is critical, but exhibits short-ranged equal-time correlations of {\it all} local operators. However, by direct calculation we find that on-site operators have power-law decay in time, e.g., $\langle \sigma^z(0)\sigma^z(\tau) \rangle \sim 1/\tau$.
This is the first microscopic model exhibiting local quantum criticality \cite{Si01}
without dissipative baths.

We acknowledge insightful discussions with M. Boninsegni, M. Freedman and M. Hastings. 
Our numerical simulations were based on the ALPS libraries \cite{ALPS}.  


\end{document}